\documentclass[sn-mathphys,Numbered]{sn-jnl}

\usepackage{graphicx}%
\usepackage{multirow}%
\usepackage{amsmath,amssymb,amsfonts}%
\usepackage{amsthm}%
\usepackage{mathrsfs}%
\usepackage[title]{appendix}%
\usepackage{xcolor}%
\usepackage{textcomp}%
\usepackage{manyfoot}%
\usepackage{booktabs}%
\usepackage{algorithm}%
\usepackage{algorithmicx}%
\usepackage{algpseudocode}%
\usepackage{listings}%

\theoremstyle{thmstyleone}%
\theoremstyle{thmstyletwo}%

\theoremstyle{thmstylethree}%

\begin{document}

\title[Article Title]{Progress toward detection of individual TLS in nanomechanical resonators}


\author[1]{\fnm{Richard} \sur{Pedurand}}

\author[1]{\fnm{Ilya} \sur{Golokolenov}}

\author[2]{\fnm{Mika} \sur{Sillanp{\"a}{\"a}}}

\author[2]{\fnm{Laure} \sur{Mercier de L{\'e}pinay}}

\author[1]{\fnm{Eddy} \sur{Collin}}

\author*[1]{\fnm{Andrew} \sur{Fefferman}}\email{andrew.fefferman@neel.cnrs.fr}

\affil[1]{\orgname{Univ. Grenoble Alpes, CNRS, Grenoble INP, Institut N{\'e}el}, \orgaddress{\country{France}}}

\affil[2]{\orgdiv{Department of Applied Physics}, \orgname{Aalto University}, \orgaddress{\country{Finland}}}


\abstract{The low temperature properties of amorphous solids are usually explained in terms of atomic-scale tunneling two level systems (TLS). For almost 20 years, individual TLS have been probed in insulating layers of superconducting quantum circuits. Detecting individual TLS in mechanical systems has been proposed but not definitively demonstrated. We describe an optomechanical system that is appropriate for this goal and describe our progress toward achieving it. In particular, we show that the expected coupling between the mechanical mode and a resonant TLS is strong enough for high visibility of a TLS given the linewidth of the mechanical mode. Furthermore, the electronic noise level of our measurement system is low enough and the anomalous force noise observed in other nanomechanical devices is absent.}

\maketitle

\section{Introduction}
At temperatures below 1 K, the mechanical, thermal and dielectric
properties of glass are dominated by low energy excitations (LEE) that are
not present in crystals. These LEE were discovered when measurements revealed that
the heat capacity of glass below 1 kelvin is much greater than that of
crystals and has a nearly linear temperature dependence, compared with the
cubic temperature dependence observed in insulating crystals \cite{Zeller71}. LEE are generally not impurities and are intrinsic to the glass since the same characteristic low temperature behaviour is observed even in very pure glass specimens \cite{Phillips87}.
According to the tunneling model the LEE are two level systems (TLS) formed by
atoms tunneling between nearby equilibria in the disordered lattice of the
glass \cite{Anderson72,Phillips72}. Under the assumption of a broad distribution of TLS energy splittings, the predictions of the standard tunneling model, in which the TLS do not interact with each other, are in good agreement with measurements of the mechanical properties of bulk insulating amorphous solids, except at the lowest temperatures \cite{Fefferman08}. The dependence of mechanical properties on the size and dimensionality of the sample yields information about the characteristic range of interactions between TLS \cite{White95}. Furthermore, the use of small amorphous samples allows one to probe individual TLS and test theoretical predictions about their behavior. A microscopic test of the tunneling model is particularly important given the competing explanations for the low energy excitations \cite{Agarwal13}.

Advances in quantum electronics have made it possible to probe individual TLS. In these impressive works, individual TLSs in the oxide barrier of Josephson junctions were probed electromagnetically \cite{Simmonds04,Martinis05,Neeley08,Lisenfeld10,Shalibo10,Sun10,Lisenfeld15}. Grabovskij et al. even studied the effect of a static applied strain on the electromagnetic response of individual TLS \cite{Grabovskij12}. Progress in this field was reviewed in \cite{Mueller19}. TLS also significantly contribute to the decoherence of superconducting qubits \cite{Klimov18,Arute19}. So far, less experimental work has concerned purely mechanical measurements of individual TLS. Such measurements are essential because, unlike in mechanical measurements, the density of TLSs inferred from dielectric measurements depends on impurity concentration \cite{Pohl02}. Thus it is not clear that dielectric and mechanical measurements are probing the same TLSs. This point was supported by combined dielectric and mechanical measurements on glass \cite{Natelson98}. Measurements of individual TLS can be carried out nanomechanically.

Nanomechanical resonators have important applications. For example, they can serve as sensitive detectors and test theories of quantum mechanics and gravity. These applications require low mechanical damping. In many cases, clamping losses are negligible, so that LEE dominate the damping of nanomechanical resonators in the low temperature limit. Therefore, it is essential to understand the behavior of the LEE in nanomechanical systems to find ways to improve their performance.

Glass nanomechanical resonators also allow tests of the tunneling model \cite{Hauer18}. Recently, Maillet et al. demonstrated agreement between measurements of 100 nm thick, high stress SiN strings covered by aluminum and predictions of the tunneling model \cite{Maillet23}. The model had to be modified relative to the standard one due to the effect of built-in stress and because the dominant thermal phonon wavelength was greater than the lateral dimensions of the string. When the superconductivity of the aluminum layer was destroyed by applying a magnetic field, the dissipation became higher, indicating that at least some of the TLS were subject to electron-driven relaxation. In contrast, a dependence of mechanical damping on superconducting state was not observed in devices made only of aluminum \cite{Kamppinen22}, indicating that further work is required for a comprehensive understanding of the effects of electron-driven TLS relaxation.

Theoretical proposals have revealed the possibility of using nanomechanical resonators to probe individual TLS \cite{Remus09,Ramos13}. The starting point of these proposals is the spectral density of the TLS and the strength of the coupling between TLS and phonons. Both of these quantities have been determined in measurements of bulk amorphous samples with large ensembles of TLS. The calculations demonstrate that, for nanometer scale samples, the number of TLS that have transition frequencies close to the mechanical resonance ranges from a few to less than one, on average. The strong sensitivity of the TLS transition frequency to static strain opens the possibility to tune TLS into resonance with the mechanical mode. A resonant TLS can be detected because the coupling between the mechanical mode and the TLS is expected to induce a splitting in the noise spectrum of the mechanical mode. The width of the splitting depends on the strength of the coupling between the TLS and the mechanical mode. However, as demonstrated by \cite{Ramos13}, the phonon occupation of the mechanical mode must be $\lesssim 1$. At higher phonon occupations this splitting is smeared and eventually disappears.

Bozkurt et al. claimed to observe signatures of individual TLS in measurements of nanomechanical resonators \cite{Bozkurt23}. In particular, a bias voltage was used to tune the transition frequencies of the TLS. Fluctuations in the mechanical time constant extracted from ringdown measurements as well as in the mechanical coherence time were observed as a function of bias voltage. This dependence on bias voltage was considered indicative of interaction between the mechanical mode and TLS defects. Furthermore, the resonant frequency of the mechanical mode varied as a function of bias voltage. However, the observed dependence of mechanical decay time, mechanical coherence time and mechanical frequency on bias voltage was not compared with predictions of theory. Since the phonon number was in the range 500-3000 in these driven experiments, the situation is different from the experiment proposed in the theoretical work \cite{Ramos13} where the signature of individual TLS only appears when the phonon number is $\lesssim 1$. Hauer et al. calculated that in their own experiment they may have achieved coupling of their mechanical modes to, on average, less than an individual thermally active defect \cite{Hauer18}. They proposed that the flattening of the mechanical damping rate below 100 mK is an observable consequence of the number of coupled TLS decreasing so that the TLS no longer acts as a bath. However, they also point out an alternative explanation for the flattening in terms of measurement-induced heating. Therefore, to our knowledge, no experiment has demonstrated definitive signatures of individual TLS coupled to nanomechanical resonators.

\section{Theory}

We plan to carry out an experiment similar to the one proposed in \cite{Ramos13}. Our optomechanical device consists of an aluminum drum vibrating at 15 MHz that is strongly coupled to a 6 GHz microwave cavity. The signature of a TLS resonant with the mechanical mode in its ground state should be easily visible according to the following calculation. We first need to determine the spectral density of TLS and their coupling to phonons in Al. An estimate can be obtained from measurements of 1 $\mu$m thick Al films \cite{Fefferman10}. In that work, the predictions of the tunneling model were fitted to the measured temperature dependence of mechanical dissipation and the change in sound speed of the Al films. The fitting parameters were the prefactor of the single phonon relaxation rate $a=4.8\times10^8$ K$^{-3}$s$^{-1}$ and the tunneling strength $C=1.3\times10^{-4}$. Assuming an average sound speed $v=5$ km/sec, this implies an average TLS/phonon coupling strength $\gamma=3.8$ eV and a TLS spectral density $P_0/\Delta_0$, where $P_0=2.3\times10^{43}$ J$^{-1}$m$^{-3}$ and $\Delta_0$ is the tunneling amplitude. An overview of the standard tunneling model and definitions of these quantities are given in \cite{Fefferman10}.

The phonon-driven relaxation rate of a symmetric TLS is $\tau^{-1}_m=a(E/k_B)^3/\tanh(E/2k_BT)$. Asymmetric TLS of the same energy $E$ relax more slowly. Phonon-driven relaxation is dominant down to 10 mK, while interactions between TLS may significantly contribute to TLS relaxation at lower temperatures \cite{Fefferman08}. We therefore evaluate the relaxation rate of a symmetric TLS with $E=(1$ mK$)k_B$ at $T=10$ mK to obtain an upper limit $\tau^{-1}_m=10$s$^{-1}$ on its relaxation rate at 1 mK. Thus a TLS that is inside our drum, which has been cooled to its ground state around 1 mK, and is nearly resonant with the 15 MHz mechanical mode should relax no faster that this rate.

Interactions between TLS also lead to spectral diffusion that explains dephasing in echo experiments \cite{Phillips87} and fluctuations in the energies of individual TLS probed with qubits \cite{Klimov18}. The elastic or electric field at the location of a given TLS $i$ changes when the configuration of TLS in its environment changes, and consequently the energy of TLS $i$ is perturbed. Only TLS in the environment of TLS $i$ with energies comparable to or less than the temperature thermally fluctuate and contribute to the energy fluctuations of TLS $i$. Along with the TLS spectral density, this yields the average distance between the fluctuating TLS and consequently the magnitude of the energy fluctuations of TLS $i$: $\Delta E_0= c C k_BT \Delta/E$, where $c$ is a constant of order unity \cite{Phillips87}. Here we have used $C=P_0\gamma^2/\rho v^2$ where $\rho$ is the mass density of the aluminum \cite{Fefferman10}. To obtain an upper limit on $\Delta E_0$, we set $\Delta=E$, insert the tunneling strength $C=1.3\times10^{-4}$ for TLS in aluminum and set $T=1$ mK, yielding $\Delta E_0^{max}/h =$(3 kHz)$c$. Klimov \emph{et al.} observed much larger TLS energy fluctuations on the scale of MHz \cite{Klimov18}. However, in contrast to the elastic coupling between TLS that we considered to estimate $\Delta E_0^{max}$,  interacting electric dipoles and ten times higher fluctuator densities than typically quoted for bulk dielectrics were required to explain the fluctuations observed in that work. The TLS we intend to probe in our nanomechanical resonator need not have dipole moments and, as mentioned in the introduction, may be different from the ones probed in electric measurements. Thus the results of Klimov \emph{et al.} may not be applicable to the case of individual TLS in mechanical resonators, which has been the object of relatively little experimental work so far.

We can now estimate the strength of the coupling $\lambda$ between the mechanical mode and a resonant TLS. This in turn determines the size of the splitting in the levels of the Jaynes-Cummings ladder formed by the mechanical mode and the TLS. Ramos \textit{et al}. calculate \cite{Ramos13}
\begin{equation} \label{lambda}
    \lambda=\frac{\gamma}{\hbar}\frac{\Delta_0}{E}S_{zpf}
\end{equation}
where $E$ is the TLS energy splitting, $S_{zpf}=\sqrt{\hbar\omega_m/2YV_m}$ is the zero-point strain in the drum, $\omega_m$ is the angular mechanical frequency, $Y=70$ GPa is the Young's modulus of the Al and $V_m$ is the mode volume. We approximate the mode volume as the actual volume of the 15 $\mu$m diameter, 100 nm thick drum. We find $S_{zpf}=6.3\times10^{-11}$ and $\lambda/\left(2 \pi\right)=58$ kHz. We estimate the average spacing between the TLS transition frequencies near $\omega_m$ using \cite{Remus09}
\begin{equation}
    \frac{\delta E}{2\pi\hbar}=\left\{\pi\hbar V_mP_0\ln\left[\frac{1+\sqrt{1-(E_{min}/\hbar\omega_m)^2}}{1-\sqrt{1-(E_{min}/\hbar\omega_m)^2}}\right]\right\}^{-1}
\end{equation}
where $E_{min}$ is the low-energy cutoff in the TLS distribution. No evidence for such a low energy cutoff was observed down to 1 mK \cite{Fefferman08}. We estimate $E_{min}/k_B=100~\mu$K and note the weak dependence of the result on this quantity. This yields a TLS frequency spacing of $\delta E/2\pi\hbar=1.4 $ MHz. Therefore, it is unlikely that a TLS will be initially within the resonant bandwidth $\lambda/\left(2 \pi\right)=58$ kHz. However, only a modest strain is required to tune a TLS into resonance with the mechanical mode. For TLS with asymmetry $\Delta<<\Delta_0$ (so that $\lambda$ is not reduced too much, Eq. \ref{lambda}), the strain required is $\delta E\Delta_0/(\Delta \gamma)=2\times10^{-9}\Delta_0/\Delta$.

\section{Experiment}

Integrating the Al drum into a microwave cavity should allow us to detect individual TLS inside the drum. In this case, motion of the drum modulates the resonance frequency of the microwave cavity, leading to optomechanical coupling \cite{Aspelmeyer14}. The fluctuations in the deflection of the drum are then imprinted on the output spectrum of the microwave cavity when the cavity is pumped. At the same time, dynamical backaction leads to optomechanical damping of the drum that increases with pump power \cite{Aspelmeyer14}. We operate in the limit of negligible optomechanical damping of the mechanical mode, so that, in the absence of a resonant TLS, the width of the mechanical peak in the spectrum corresponds to the intrinsic damping rate of the mechanics. In the case where the bath is cold enough so that the drum and TLS are in their ground states, a fine structure in the mechanical peak is predicted to appear \cite{Ramos13}. The size of the splitting in the peak due to the resonant TLS is given by the coupling between the TLS and the mechanical mode $\lambda/\left(2 \pi\right)\approx58$ kHz. Our drum has an intrinsic linewidth of 400 Hz in the low temperature limit \cite{Cattiaux21}. Furthermore, as calculated above, the upper limits of the TLS relaxation rate and energy fluctuations are respectively $\tau^{-1}_m=10$s$^{-1}$ and $\Delta E_0^{max}/h \approx 3$ kHz. Thus the expected splitting is much greater than the mechanical and TLS linewidths, which should make the signature of an individual TLS visible.

Existing optomechanical devices, including our own, are therefore apparently suitable for observing individual TLS. However, a successful experiment of the type proposed here requires passive cooling to the mechanical ground state. If the temperature is too high, the fine structure of the microwave output spectrum is smeared \cite{Ramos13}. Furthermore, the device should be free of anomalous force noise acting on the mechanics so that the signature of a TLS can be clearly distinguished. This force noise is well known in the field of low temperature nanomechanics and was first discussed explicitly in \cite{Zhou19}. Its origin remains unclear, but it is known to be more severe in strings than in drums. Passive ground state cooling of our drum was reported in \cite{Cattiaux21}. However, the noise floor of the detection system was not optimized, requiring averages on the timescale of days. TLS drifting out of resonance with the mechanical mode on this timescale might not have been observable. We have therefore greatly improved our noise floor as a step toward observing individual TLS. This has also allowed us to confirm that our drum is not subject to the large anomalous force noise that afflicted the vibrating string studied in \cite{Zhou19}.

\begin{figure}
\includegraphics[width=\textwidth]{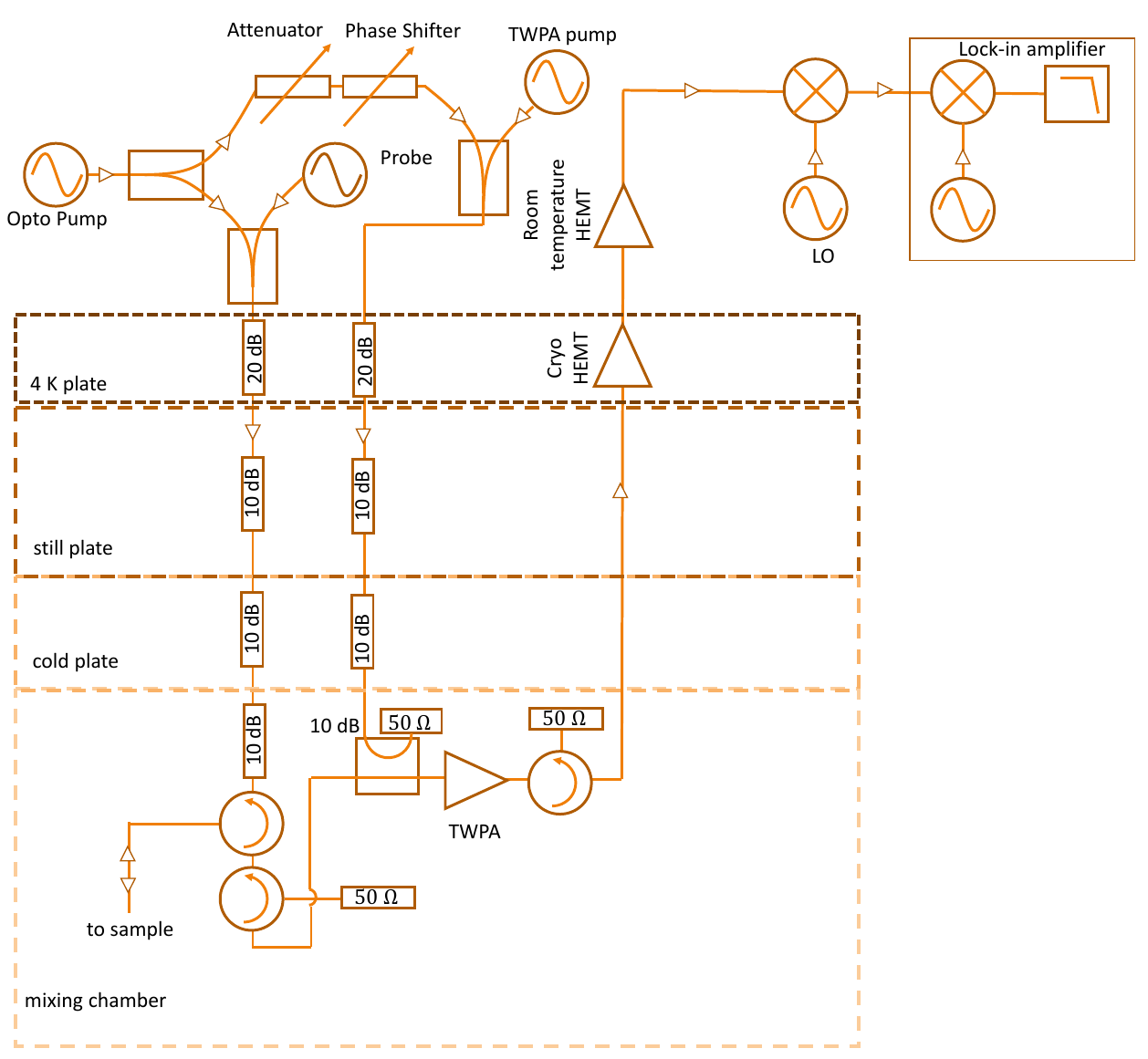}
\caption{Microwave circuit used to measure the vibrations of the drum. One drive line is used for the optomechanical pump and the probe. The other drive line is used for the TWPA pump and to cancel the optomechanical pump that is reflected from the sample. A directional coupler allows us to combine the signal reflected from the sample with the opposition signal and the TWPA pump with small attenuation of the signal from the sample. The circulators are used to isolate the sample and TWPA from spurious signals. This circuit was installed on a cryostat containing a liquid helium bath.}
\label{fig:cct}
\end{figure}

The microwave circuit diagram used to measure the drum is shown in Fig. \ref{fig:cct}. The placement and values of the attenuators on the drive lines were chosen to minimize the thermal noise power reaching the sample while avoiding excessive heat loads on the cryostat due to the microwave pumps. A travelling wave parametric amplifier (TWPA) was used in order to maximize the signal to noise ratio. The circuit is very similar to the one described in \cite{Golokolenov23b}, where more details are given. In contrast with that work, the circuit was installed on a cryostat coupled to a liquid helium bath. The cryostat is described in the supplemental information of \cite{Cattiaux21}. Because the microwave resonance in the present work was at a slightly different frequency from the one in \cite{Golokolenov23b}, we achieved a TWPA gain as high as 18 dB at 20 mK compared with 10 dB in \cite{Golokolenov23b}. We measured the amplification of a probe tone with the TWPA turned on and off and found a 14 dB enhancement of the signal to noise ratio with the TWPA on in the present work. The noise temperature of the cryogenic HEMT that followed the TWPA was 1.6 K at 5 GHz. We verified experimentally that the system white noise is dominated by the cryogenic components and that it increased by as much as 50\% when turning on the TWPA. Since the noise at the TWPA input is not dominating the system noise, an increase in the TWPA gain would improve the signal to noise ratio above its already impressive level. As discussed below, the system noise level is already low enough so that TLS may be detected upon cooling the mechanics to the ground state.

\section{Results}
In the present work we used our increased sensitivity relative to \cite{Cattiaux21} to verify the expected linear temperature dependence of the thermomechanical noise power to a higher precision than in \cite{Cattiaux21}. This demonstrates that the mechanical mode is in thermal equilibrium with the cryostat, since the thermomechanical noise power would stop decreasing with cryostat temperature in the case of thermal decoupling. As explained above, cooling the device is important for avoiding smearing of the fine structure that signifies coupling of the mechanical mode to an individual TLS. In cavity optomechanics, the thermomechanical noise is imprinted on the output spectrum of the pumped cavity \cite{Aspelmeyer14}. At high pump powers, dynamical backaction increases or decreases the effective damping rate of the mechanics depending on the detuning of the pump relative to the microwave resonance. Here, we aimed to measure the intrinsic properties of the mechanics without having to extrapolate the measurements to lower pump powers. The weak pump was detuned above the microwave cavity resonance by an amount equal to the mechanical resonance frequency (“blue pumping”) so that the lower mechanical sideband of the pump was aligned with the microwave resonance.

\begin{figure}
\includegraphics[width=\textwidth]{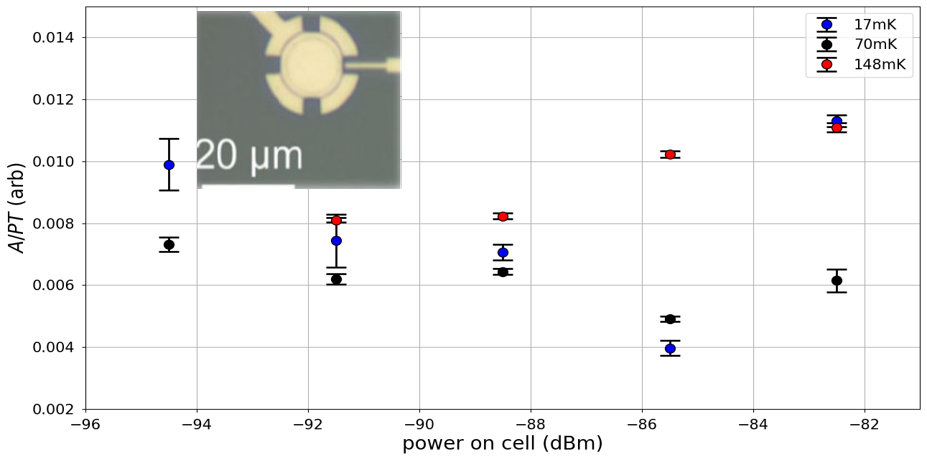}
\caption{Mechanical sideband area normalized by the product of pump power and temperature. The lack of dependence of this ratio on pump power and temperature demonstrates thermal equilibrium between the mechanical mode and the cryostat. Thousands of power spectra were averaged to obtain these data, and the error bars show the standard deviation of the mean area. The total required data acquisition time was five hours for each of the points in this figure, except for the ones at 17 mK and -94.5 and -91.5 dBm pump power, which required 16 and 47 hours, respectively. The drum studied in the present work is the same as the one used by Cattiaux \emph{et al.}, and a picture of a device from the same batch is shown in the inset \cite{Cattiaux21}.}
\label{fig:area}
\end{figure}

Figure \ref{fig:area} shows the area of this sideband, normalized by the pump power and the cryostat temperature (A/PT), as a function of pump power and cryostat temperature. At these low pump powers, where dynamical backaction on the mechanics is negligible, and with constant internal and external microwave cavity losses, we observed the quantity A/PT to be independent of temperature and pump power, as expected \cite{Cattiaux21}. Our pump powers were low enough so that we also did not observe spurious population of the mechanical mode (the “technical heating” discussed in \cite{Cattiaux21}). The low electronic noise of our measurement system allowed us to measure the mechanical noise with very low power. The minimum pump power in Fig. \ref{fig:area} corresponds to only $n_{cav}=17$ photons stored in the cavity \cite{Cattiaux21}. Such low values of $n_{cav}$ were also achieved in the seminal work by Teufel \emph{et al.}, where the optomechanical damping under red pumping was approximately equal to the intrinsic damping of 30 Hz at $n_{cav}=20$ \cite{Teufel11a}.

\begin{figure}
\includegraphics[width=\textwidth]{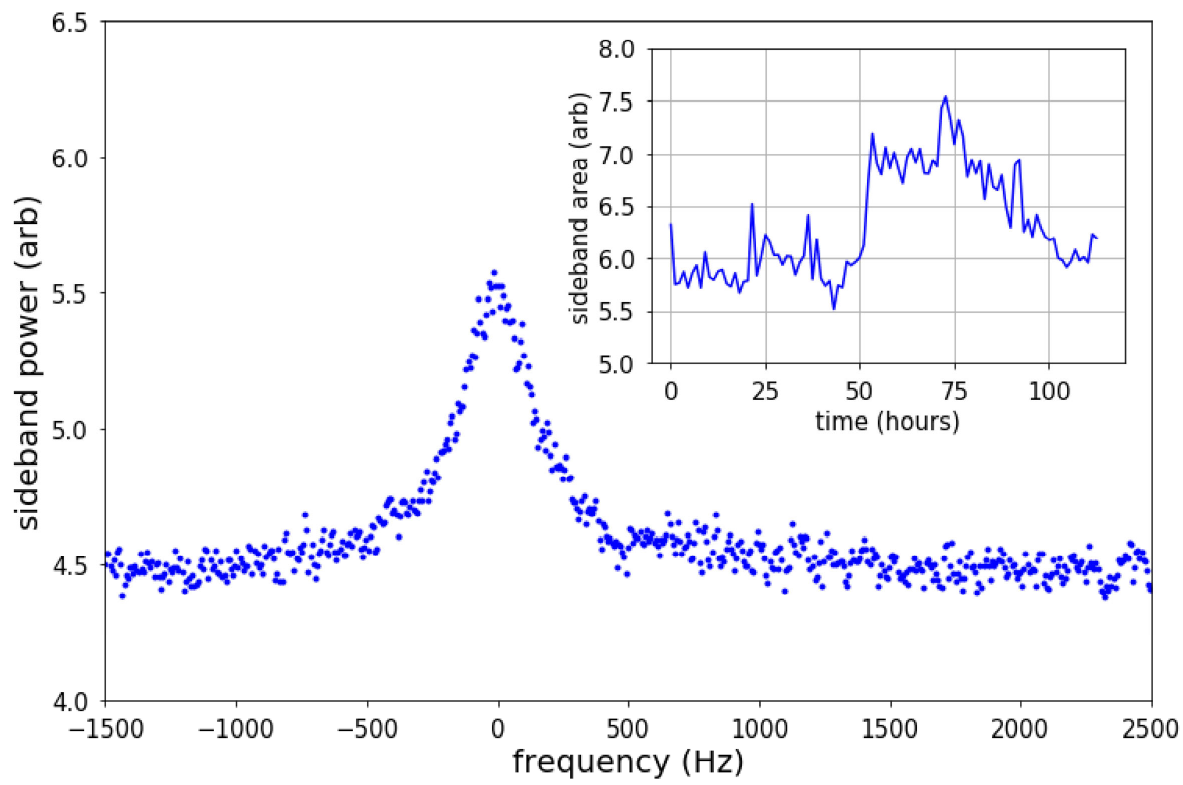}
\caption{Mechanical sideband at 17 mK under weak blue pumping with a power of -83 dBm at the cell. The spectrum was obtained after averaging for approximately one hour. Inset: the variation of the sideband area with time is very small compared with the variations by a factor of up to 10$^5$ observed in a SiN/Al string \cite{Zhou19,KumarThesis}. Movies showing the time evolution of the spectrum with different averaging times are in the supplemental material.}
\label{fig:stability}
\end{figure}

As discussed above, some optomechanical devices are plagued by an anomalous force noise that interferes with the measurement of noise temperature and could interfere with the observation of individual TLS as well. Figure \ref{fig:stability} demonstrates the absence of such "spikes" in the mechanical noise spectrum in this device at 17 mK over at least a 100 hour timescale. This rules out aluminum as an intrinsic source of the anomalous force noise. Potential explanations for the spikes that appeared in measurements of an Al/SiN string are discussed in detail in \cite{KumarThesis}.

At 1 mK, the mean thermal phonon occupation of our 15 MHz mode is one. As explained above, the signature of a TLS resonant with the mechanical mode is expected to be visible at this occupation number. Since the area of the Lorentzian anti-Stokes sideband is proportional to the phonon occupation, and the mechanical linewidth of our device is temperature independent below 20 mK \cite{Cattiaux21}, the amplitude of the peak will decrease by a factor of 24 at 1 mK compared to the peak in Fig. \ref{fig:stability}. We do not expect any change in the electronic noise level in this temperature range. The mechanical peak amplitude will then be difficult to distinguish from the electrical background noise with one hour averaging time, so the averaging time will have to be approximately one day at the same pump power. We can use the same -83 dBm pump power because it does not cause significant technical heating at 0.75 mK \cite{Cattiaux21}. The fluctuations in the TLS energy are expected to be about 3 kHz for arbitrarily long averaging time (see above) and therefore much less than the coupling bandwidth $\lambda/2\pi=58$ kHz. We therefore expect the capability to extend the averaging time far beyond one hour to achieve a high signal to noise ratio.

\section{Conclusion}
The standard tunneling model is often used to explain the temperature dependence of mechanical, electric and thermal properties in the millikelvin range. For many years, the model was used with limited microscopic justification because of the simplicity of the underlying assumptions and the generally good agreement with experiments. Recently developed nanofabrication techniques and exquisitely sensitive measurement systems have enabled measurements of individual TLS in qubits, allowing microscopic tests of the tunneling model. These experiments have upheld the validity of the tunnling model, but it is not clear that the TLS probed in these experiments are the same as the ones responsible for mechanical properties of glass. We have summarized leading theoretical approaches and experimental progress toward probing individual TLS in mechanical resonators. We believe that no experiment has demonstrated definitive signatures of individual TLS coupled to nanomechanical resonators and that our optomechanical device is a promising system for doing so. We have demonstrated great improvement in the electrical noise background of our measurement, and the lack of anomalous features in the mechanical noise spectrum is promising for detection of individual TLS. Our estimates of the coupling strength between our mechanical mode and resonant TLS indicate that individual TLS can be observed at temperatures within the range of our cryostat. Future work will include controlling the strain in the drum by applying a bias voltage and extending our measurements to the quantum ground state again, this time with the improved electrical noise level.

\section{Acknowledgements}
We acknowledge support from the European Research Council under StG UNIGLASS Grant No. 714692 (A.F.) The research leading to these results has received funding from the European Union’s Horizon 2020 Research and Innovation program, under Grant No. 824109, the European Microkelvin Platform. We acknowledge the facilities and technical support of Otaniemi research infrastructure for Micro and Nanotechnologies (OtaNano) (M.S.). This work was supported by the Academy of Finland (contracts 352189, 352932, and 336810) (M.S.), and by the European Research Council (contract 101019712) (M.S.), and by the QuantERA II Programme (contract 13352189) (M.S.).


\end{document}